# Absence of superconductivity and density-wave transition in ambient-pressure tetragonal La$_4$Ni$_3$O$_{10}$

Mengzhu Shi[1,2,†], Yikang Li[1,2,†], Yuxing Wang[3], Di Peng[6], Shaohua Yang[8], Houpu Li[1,2], Kaibao Fan[1,2], Kun Jiang[3], Junfeng He[1,5], Qiaoshi Zeng[6,7], Dongsheng Song[8], Binghui Ge[8], Ziji Xiang[2,5], Zhenyu Wang[1,5], Jianjun Ying[1,5], Tao Wu[1,2,4,5,*] and Xianhui Chen[1,2,4,5,*]

1. CAS Key Laboratory of Strongly-coupled Quantum Matter Physics, Department of Physics, University of Science and Technology of China, Hefei, Anhui 230026, China

2. Hefei National Laboratory for Physical Sciences at the Microscale, University of Science and Technology of China, Hefei, Anhui 230026, China

3. Institute of Physics, Chinese Academy of Sciences, and Beijing National Laboratory for Condensed Matter Physics, Beijing, 100190, China

4. Collaborative Innovation Center of Advanced Microstructures, Nanjing University, Nanjing 210093, China

5. Hefei National Laboratory, University of Science and Technology of China, Hefei 230088, China

6. Shanghai Key Laboratory of Material Frontiers Research in Extreme Environments (MFree), Institute for Shanghai Advanced Research in Physical Sciences (SHARPS), Shanghai, China

7. Center for High Pressure Science and Technology Advanced Research, Shanghai, China

8. Information Materials and Intelligent Sensing Laboratory of Anhui Province, Anhui Key Laboratory of Magnetic Functional Materials and Devices, Institutes of Physical Science and Information Technology, Anhui University, Hefei, China

†These authors contributed equally to this work
*Correspondence to: wutao@ustc.edu.cn, chenxh@ustc.edu.cn

**The recent discovery of superconductivity in La$_3$Ni$_2$O$_7$ and La$_4$Ni$_3$O$_{10}$ under high pressure stimulates intensive research interests. These nickelates crystallize in an orthogonal/monoclinic structure with tilted NiO$_6$ octahedra at ambient pressure and enter a density-wave-like phase at low temperatures. The application of pressure suppresses the octahedral tilting and triggers a transition to tetragonal structure (*I4/mmm*), which is believed to be a key prerequisite for the emergence of superconducting state. Here, by developing a high oxidative environment growth**

**technology, we report the first tetragonal nickelates $La_4Ni_3O_{10}$ microcrystals without octahedral tilting at ambient pressure. In tetragonal $La_4Ni_3O_{10}$, transport measurements find that both density-wave and superconducting transitions are absent up to 160 GPa, indicating a robust tetragonal metallic ground state. Density functional theory calculations reveal that the band structure of ambient-pressure tetragonal $La_4Ni_3O_{10}$ involves more $d_{z^2}$ orbital contribution to the Fermi surface, compared to the monoclinic phase or the high-pressure superconducting tetragonal phase. The concurrent absence of density-wave state and high-pressure superconductivity in our ambient-pressure tetragonal crystals of $La_4Ni_3O_{10}$ suggests an underlying correlation between these two orders. It suggests that the tetragonal structure is not necessary, while the density-wave state is crucial for the superconductivity in nickelates. Our findings impose important constraints on the mechanism of pressure-induced superconductivity in nickelates and sheds new light on exploring ambient pressure high-temperature Ni-based superconductors.**

Since the milestone discovery of high-temperature superconductivity in $La_{2-x}Ba_xCuO_4$ [1], whether high-temperature superconductivity could be achieved in its nickelate analog has become an outstanding question for the community [2–10]. In 2019, the discovery of superconductivity in infinite-layer nickelate $Nd_{1-x}Sr_xNiO_2$ thin film, which shares a similar crystalline structure as the infinite cuprate superconductor $CaCuO_2$ [12] and shows a superconducting transition temperature ($T_c$) ranging from 9 to 15 K [11], opens a fresh avenue for exploring the high-$T_c$ puzzle. Very recently, nickelate superconductor family has been successfully extended to the Ruddlesden-Popper (RP) phases $La_{n+1}Ni_nO_{3n+1}$ with n = 2 [13] and n = 3 [14–16] and the maximum of $T_c$ determined by transport measurement on $La_3Ni_2O_7$ is beyond the liquid nitrogen temperature. However, different from that in infinite-layer nickelates $Nd_{1-x}Sr_xNiO_2$ thin film, the realization of superconductivity in the RP phase $La_{n+1}Ni_nO_{3n+1}$ requires a high-pressure ($P$) environment [13], which greatly hinders the spectroscopic and thermodynamic studies of the superconducting state and thus the pairing symmetry. Up to now, the experimental study of the high-pressure superconductivity in RP phase $La_{n+1}Ni_nO_{3n+1}$ is still rather limited and the superconducting mechanism is under hot debate [17]. The exploration of superconductivity at ambient pressure in the RP phase $La_{n+1}Ni_nO_{3n+1}$ is one of the most important directions to solve above dilemma.

In both La$_3$Ni$_2$O$_7$ and La$_4$Ni$_3$O$_{10}$, the NiO$_6$ octahedron is tilted from the *c* axis at ambient pressure [13–16], leading to an orthogonal or monoclinic distortion of tetragonal structure (Fig. 1a-1d). The bond angle of Ni-O-Ni between adjacent NiO$_6$ octahedrons is 168° (167°) in La$_3$Ni$_2$O$_7$ (La$_4$Ni$_3$O$_{10}$) which is thought to be critical for the interlayer coupling between NiO planes [13]. These distorted structures host spin and/or charge orders as implied by density-wave-like transitions in resistivity [18–24]. In La$_4$Ni$_3$O$_{10}$, neutron scattering and X-ray measurements indicate that the density-wave transition at ~ 136 K involves a simultaneous development of spin-density-wave (SDW) order and charge-density-wave (CDW) order [25]. In La$_3$Ni$_2$O$_7$, although a similar SDW order has been confirmed by recent spectroscopy measurements below ~ 150 K [26–29], the existence of CDW order is still elusive [17]. With increasing pressure, the volume of unit cell shrinks remarkably, making the titled NiO$_6$ octahedron unstable [13–16]. In orthogonal La$_3$Ni$_2$O$_7$, a pressure-induced structure transition from Amam to Fmmm has been observed by recent X-ray diffraction experiment under high pressure [30]. Moreover, more X-ray diffraction experiments under high pressure further demonstrate that, when the superconductivity appears with increasing pressure, a first-order structural transition into the tetragonal phase (*I4/mmm*) occurs almost concurrently in both pressurized La$_3$Ni$_2$O$_7$ and La$_4$Ni$_3$O$_{10}$ [15,16,30,31] (see Fig. 1e). Moreover, the density-wave transition completely disappears in the high-pressure tetragonal structure [13–16,32,33]. As such, the tetragonal structure (*I4/mmm*) is conjectured to be a prerequisite for the emergence of superconductivity [17], whereas the relation between the density-wave order and superconductivity remains unclear [29–31]. Here, with high-oxidative-environment growth technology, we successfully synthesize tetragonal La$_4$Ni$_3$O$_{10}$ microcrystals without octahedral tilting at ambient pressure. Both density-wave and superconducting transitions are absent up to 160 GPa in the tetragonal La$_4$Ni$_3$O$_{10}$ phase. Such unexpected absence of electronic orders in tetragonal RP phase nickelates put constraints on the theories modeling the interactions therein and provide important insights into the mechanism of pressure-induced superconductivity.

**Tetragonal La$_4$Ni$_3$O$_{10}$ at ambient pressure**

Fig. 1c and 1d show the structures of monoclinic La$_4$Ni$_3$O$_{10}$ (ambient pressure) and tetragonal superconducting La$_4$Ni$_3$O$_{10}$ (high pressure phase) [15,16], respectively. Both structures consist of a (LaNiO$_3$)$_3$ sub-structure which contains three-layer perovskites inter-grown with a LaO rock-salt sub-structure, forming the RP phase. For the ambient-pressure monoclinic phase, the bond angle of O-Ni-O in the NiO$_6$ octahedron is about 177° (Fig. 2a). Furthermore, the adjacent NiO$_6$ octahedron tilts towards each other. The bond angle of Ni-O-Ni between the adjacent NiO$_6$ is about 167° (Fig. 2a). Such a tilt introduces a buckling Ni-O plane in the perovskite layer. However, in the high-pressure tetragonal phase, the bond angles of O-Ni-O in the NiO$_6$ octahedron are all 180° (along *a*, *b* and *c* axis Fig. 2b). The adjacent NiO$_6$ octahedrons arrange parallel to each other, and the bond angles of Ni-O-Ni between the adjacent NiO$_6$ octahedrons are also 180° (Fig. 2b). The absence of tilting between the adjacent NiO$_6$ octahedron leads to a flat plane of the perovskite sub-structure. The temperature-pressure phase diagram of La$_4$Ni$_3$O$_{10}$ (Fig. 1e) implies that the phase boundary between the density wave state and superconductivity is close to the pressure-induced monoclinic to tetragonal structural transition. This suggests a substantial impact of the structural transition on the electronic band structure, and a tetragonal phase seems to be more favorable to the appearance of superconductivity. More importantly, such correlation between structural transition and the emergence of superconductivity has been revealed in both La$_3$Ni$_2$O$_7$ and La$_4$Ni$_3$O$_{10}$ [13-16]. As a result, this engenders a fundamental question: can the tetragonal phase, supposedly hosting superconducting ground state, be stabilized at ambient pressure? Indeed, a metastable orthorhombic phase (Bmab) has been reported to be stabilized by postgrowth rapid cooling under 20 bar oxygen pressure [34]. This might give a clue that the growth of La$_4$Ni$_3$O$_{10}$ with higher structural symmetry could be favored in a strong oxidative atmosphere. We have adopted an oxidative molten salt method under oxygen pressure to synthesize La$_4$Ni$_3$O$_{10}$ crystal (details see method). The major products are the ambient-pressure tetragonal La$_4$Ni$_3$O$_{10}$ microcrystals.

The structure of the as-grown La$_4$Ni$_3$O$_{10}$ microcrystal grown under flowing oxygen environment is solved by the refinement of single crystal X-ray diffraction (SC-XRD) data. The reciprocal lattice data along *a**, *b** and *c** axes is shown in the Extended Data Fig. 1 which satisfies a *I4/mmm* space group and thus corroborates the tetragonal symmetry. The details of crystal structure solved from the SC-XRD at 300 K is shown in Table 1. Furthermore, the selected area electron diffraction (SAED) data along the along the [100] and [001] axis shown in the Extended Data Fig. 2 also confirms the tetragonal structure solved by the SC-XRD data. For comparison, the reciprocal lattice data for the monoclinic La$_4$Ni$_3$O$_{10}$ single crystal is shown in the Extended Data Fig.3 and the detail structure information is also shown in Table 1. Fig. 2a-2c summarizes the connection between the NiO$_6$ octahedron for monoclinic phase at ambient pressure ($P2_1/a$-AP), tetragonal phase under high pressure ($I4/mmm$-HP) [15,16] and tetragonal phase at ambient pressure ($I4/mmm$-AP), respectively. A general description of the connection between the $NiO_6$ octahedron in these three structures is shown in Fig. 2d. We note that site O5 only exists in the monoclinic phase, which becomes equivalent to site O2 in the tetragonal phases due to higher symmetry. The detailed bond lengths and bond angles are also provided in Fig. 2e. Considering these results, we note that the $I4/mmm$-AP phase and $I4/mmm$-HP phase is almost identical with only slightly different specific bond lengths. The bond angle in the $NiO_6$ octahedron (Ni3-O2-Ni4 and O4-Ni1-O4) and between the adjacent $NiO_6$ (Ni2-O4-Ni1) are all 180° in these two tetragonal phases. Furthermore, the powder XRD pattern for tetragonal $La_4Ni_3O_{10}$ is also collected by grinding several pieces of tetragonal single crystals under the pressure range from ambient pressure to 27.7 GPa. As shown in Fig. 2f, all diffraction peaks can be well indexed by the $I4/mmm$ tetragonal structure. For example, to enlarge the diffraction peaks between 10 and 11.5 degrees as shown in Fig. 2g, the two diffraction peaks can be well indexed by (107) and (110) for a tetragonal structural phase and no additional splitting due to the monoclinic distortion is observed in this range of $2\theta$. These results indicate that the tetragonal structure is kept with pressure at least up to 27.7 GPa. The specific lattice parameters and cell volume of the tetragonal $La_4Ni_3O_{10}$ under different pressures are shown in the Extended Data Fig. 4. In addition, it should be noted that the tetragonal structure is also kept down to 110 K under ambient pressure (see the Extended Table 1).

In Fig. 1f, we compare the temperature dependent resistance curves $R(T)$ for monoclinic $La_4Ni_3O_{10}$ (sample S1) and ambient-pressure tetragonal $La_4Ni_3O_{10}$ (sample S3). There is a density wave transition manifested by a huge hump in the resistivity at around 135 K for the monoclinic phase, which is consistent with the previous report [35]. On the other hand, the $R(T)$ of the tetragonal phase shows a metallic behavior without notable anomalies, hinting at the absence of the density wave order. Moreover, no superconductivity is observed in our tetragonal $La_4Ni_3O_{10}$ down to 0.4 K. Such metallic behavior without density wave transition has been repeated on three pieces of tetragonal samples (the Extended Data Fig. 6). To confirm the absence of density wave transition in our tetragonal sample, we have also performed magnetic torque measurement on both monoclinic and tetragonal $La_4Ni_3O_{10}$ samples (see method for details). As shown in Fig. 1g, a prominent feature around 135 K is observed in the monoclinic $La_4Ni_3O_{10}$ sample, implying the impact of the density-wave transition on magnetization, while it is absent in the tetragonal La$_4$Ni$_3$O$_{10}$ sample.

**Pressure effect in monoclinic and tetragonal La$_4$Ni$_3$O$_{10}$**

To further investigate the physical properties of La$_4$Ni$_3$O$_{10}$ with different structures, we performed high-pressure resistance measurements using liquid pressure medium (see details in method). For the monoclinic La$_4$Ni$_3$O$_{10}$ (sample S1), the resistance gradually decreases with increasing pressure, and the anomaly associated with the density wave transition can be gradually suppressed as shown in Fig. 3a. Above 20 GPa, the density wave transition is completely suppressed; a weak resistive drop occurs at ~ 3 K at 25.3 GPa, signifying a superconducting transition. With further increasing pressure, the transition becomes more pronounced and its temperature $T_c$ gradually increases; zero resistance is achieved at around 4 K at $P$ = 72 GPa (Fig. 3b). We have also measured another monoclinic La$_4$Ni$_3$O$_{10}$ (sample S2) using a solid pressure medium (see details in method), the data are presented in Extended Data Fig. 7. Superconductivity is observed above 24 GPa, with a maximum $T_c$ of up to 24 K, similar to that in sample S1. The $T$-$P$ phase diagram constructed based on these data (Fig. 1e) is consistent with previous findings [14–16].

In contrast, for the tetragonal La$_4$Ni$_3$O$_{10}$ (sample S3), no anomaly related to the density wave transition is observed at ambient pressure. The general metallic behavior of the $T$-dependent resistance persists up to above 60 GPa without indications of superconductivity (Fig. 3c and 3d). The resistance gradually decreases with increasing pressure below 14 GPa, then starts to increase as shown in Fig. 3c. Above 20 GPa, the resistance decreases again with increasing pressure and shows a weak upturn at low temperatures as shown in Fig. 3d; the nonmonotonic $P$-dependence is illustrated in Fig. 3f. We have also conducted high-pressure resistance measurements on another tetragonal La$_4$Ni$_3$O$_{10}$ (sample S4) up to 158.9 GPa, as shown in Extended Data Fig. 8. It exhibits similar behaviors as those in sample S3, and no superconductivity is observed up to the highest pressure. From our high-pressure resistance measurements, we conclude that the as-grown tetragonal La$_4$Ni$_3$O$_{10}$ exhibits strikingly different high-pressure behavior comparing to that in the monoclinic La$_4$Ni$_3$O$_{10}$, although the monoclinic structure can also be tuned to tetragonal phase at a certain pressure where the superconductivity sets in (Fig. 1e).

**DFT calculations of monoclinic and tetragonal La$_4$Ni$_3$O$_{10}$**

The band structures and Fermi surfaces (FSs) of three different $La_4Ni_3O_{10}$ crystal structures have been investigated by performing first-principles calculations based on the density functional theory (DFT). For the $P2_1/a$-AP phase and the $I4/mmm$-HP phase, the band structures and Fermi surfaces are shown in Fig. 4a, 4b and 4d, 4e, respectively. In the case of the $P2_1/a$ phase, the tilted $NiO_6$ octahedrons fold the Brillouin zone (BZ) into the dashed $\sqrt{2} \times \sqrt{2}$ BZ displayed in Fig. 4d, whereas the tetragonal phases are characterized by the typical square BZ shown in Fig. 4d. Considering the Ni 3$d$ orbitals, there are 12 and 6 $e_g$ bands for the $P2_1/a$-AP (Fig. 4a) and $I4/mmm$-HP (Fig. 4b) phases, respectively. The FSs for both phases comprise preponderant $d_{x2-y2}$ orbital component, which occasionally mixes with the $d_{z2}$ orbitals (Fig. 4d and 4e); The calculated FSs of the $P2_1/a$-AP phase (Fig. 4d) matches well with the recent ARPES results [36]. For the newly-synthesized $I4/mmm$-AP phase introduced in this work, the band structure and FSs are presented in Fig. 4c and 4f for a comparison. We can see that the most salient feature for this phase is the appearance of two additional FS pockets $\gamma, \varepsilon$ centered at the Γ point and M point, respectively. According to the orbital projection plot (Fig. 4f), these additional FSs are mainly attributed to the $d_{z2}$ orbital. Their emergence is due to the elongated $c$ axis in the $I4/mmm$-AP structure which reduces the $d_{z2}$ interlayer coupling. On the contrary, the $I4/mmm$-HP phase suffers from the compressed $c$ axis and thus the $\gamma, \varepsilon$ pockets disappear (Fig. 4e).

**Discussion**

Now we turn to discuss the possible implications of the present work on exploring the superconducting mechanism in RP-phase nickelate superconductors. In previous studies, it is broadly speculated that a key impact of high pressure on introducing superconductivity is to stabilize the tetragonal structure without octahedron tilting in both $La_3Ni_2O_7$ and $La_4Ni_3O_{10}$. Such a scenario has immediately revealed the importance of synthesizing tetragonal $La_3Ni_2O_7$ and $La_4Ni_3O_{10}$ crystals at ambient pressure. By substantially reducing the oxygen content in both compounds, the tetragonal structure is indeed achieved at ambient pressure in $La_3Ni_2O_6$ and $La_4Ni_3O_8$, respectively [24,37,38]. Previous TEM results indicate that the removal of oxygen predominantly happens in the apical oxygen sites [39]. In these oxygen-deficient tetragonal phases, an insulating transport behavior always prevails at low temperatures, which inevitably hampers the development of superconductivity

[24,40]. For our $I4/mmm$-AP La$_4$Ni$_3$O$_{10}$ samples grown in high oxygen atmosphere, although precise determination of the oxygen content is difficult due to the small sample size, the XRD refinement of crystal structure implies an oxygen content very close to the stoichiometry (see methods). By this way, the oxygen content of the $I4/mmm$-AP La$_4$Ni$_3$O$_{10}$ is found to be slightly larger than that of the $P2_1/a$-AP La$_4$Ni$_3$O$_{10}$ by ~ 0.04. Furthermore, we also used integrated differential phase contrast (iDPC) imaging techniques to accurately visualize the oxygen atoms on the tetragonal La$_4$Ni$_3$O$_{10}$. As shown in Extended Data Fig. 5, this result shows no obvious oxygen vacancy or defects. In this sense, the absence of superconducting state in the $I4/mmm$-AP La$_4$Ni$_3$O$_{10}$ samples at both ambient and high pressure can hardly be attributed to oxygen deficiency (this is also demonstrated by the metallic resistance behavior shown in Fig. 3c), but it places a direct challenge on the idea that the high-pressure superconductivity ubiquitously emerges in RP-phase nickelates once a tetragonal structure develops.

A tempting explanation is the existence of robust magnetic order in these ambient-pressure tetragonal samples. In monoclinic La$_4$Ni$_3$O$_{10}$, SDW order has already been revealed by the neutron scattering experiment [25]. Very recently, presence of a similar SDW order has been confirmed in La$_3$Ni$_2$O$_7$ by various spectroscopy probes including resonant inelastic X-ray scattering (RIXS), nuclear magnetic resonance (NMR) and muon spin rotation (μSR) experiments [26–29]. Our DFT calculation indicates that the magnetic ground state is still more favorable in the $I4/mmm$-AP phase at least in the case without applying pressure (see methods and Extended Table 2 and 3). However, our transport and magnetic torque measurement does not show any signature for the SDW transition up to 160 GPa. One possible reason is that multiple competing (and nearly-degenerate) magnetic states coexist in the $I4/mmm$ structure give rise to a frustrated paramagnetic ground state. Further spectroscopy measurements (such as NMR and μSR) are urgently needed to clarify the ground state and examine whether the absence of SDW transition and superconductivity is internally connected. In addition, the role of $d_{z^2}$ orbital on the high-pressure superconductivity in La$_3$Ni$_2$O$_7$ and La$_4$Ni$_3$O$_{10}$ is under hot debate [17]. In the tetragonal La$_4$Ni$_3$O$_{10}$ at ambient pressure, our DFT calculations indicate that the $d_{z^2}$ orbital contributes two additional FS pockets located at the Γ point and M point, respectively (Fig. 4f). With increasing pressure, these two pockets ($\gamma, \varepsilon$) disappear from the Fermi level, suggesting a Lifshitz transition. By analyzing the pressure-dependent resistance, a nonmonotonic behavior is observed around 20 GPa (see Fig. 3f), which further supports the occurrence of a Lifshitz transition. The superconducting state is rigorously absent in the *I4/mmm*-AP samples, as mentioned above, both at ambient pressure and under applied high pressure up to 160 GPa, i.e., irrespective of the presence or absence of the $d_{z^2}$-derived FS pockets. Therefore, our results do not support a critical role of $d_{z^2}$ orbital on high-pressure superconductivity. As a final note, spectroscopy measurements on the new *I4/mmm*-AP La$_4$Ni$_3$O$_{10}$ are needed in the near future, which would be undoubtedly decisive for understanding the key ingredients in the electronic states that determining the fate of superconductivity in RP-phase nickelates.

State in the Triple-Layer T'-La$_4$Ni$_3$O$_8$. *Phys. Rev. Lett.* **108**, 236403 (2012).

# Figures and Tables

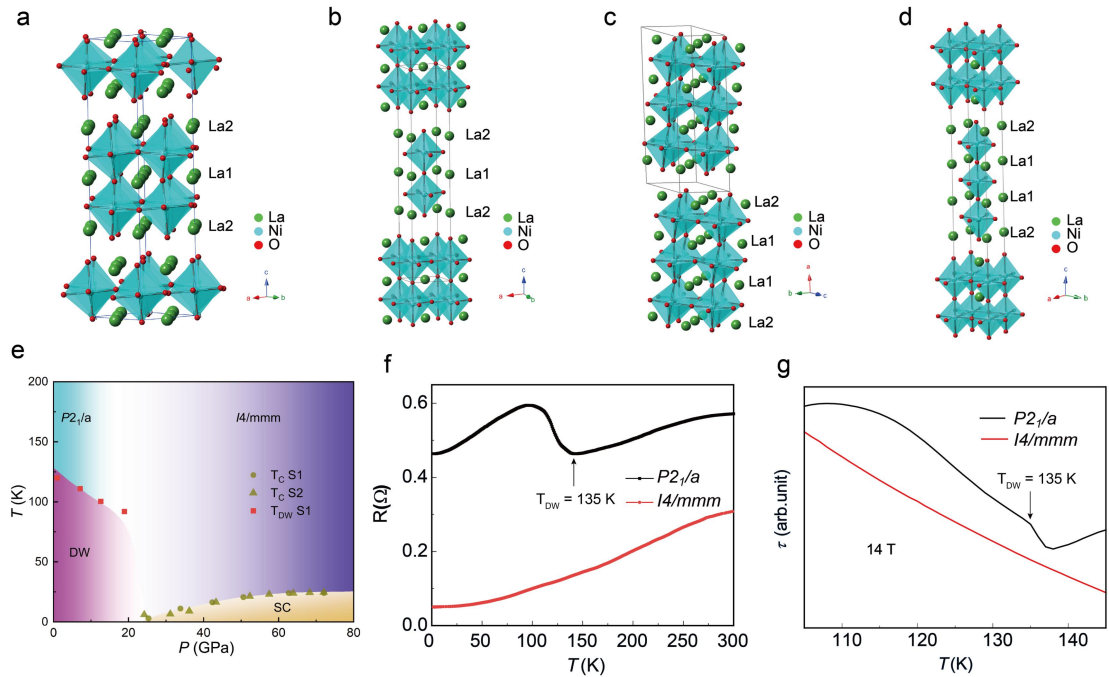

**Fig. 1. Crystal structure, phase diagram, resistance curves and magnetic torque curves of La$_4$Ni$_3$O$_{10}$.** (a) and (b), the crystal structure of orthogonal La$_3$Ni$_2$O$_7$ at ambient pressure and tetragonal La$_3$Ni$_2$O$_7$ under pressure. There is a small tilting angle between the adjacent NiO$_6$ octahedron in the orthogonal La$_3$Ni$_2$O$_7$, which disappears in the tetragonal La$_3$Ni$_2$O$_7$. (c) and (d), the crystal structure of monoclinic La$_4$Ni$_3$O$_{10}$ at ambient pressure and tetragonal La$_4$Ni$_3$O$_{10}$ under pressure. A small tilting angle between the adjacent NiO$_6$ octahedron also exists in the monoclinic La$_4$Ni$_3$O$_{10}$ and disappears in the tetragonal La$_4$Ni$_3$O$_{10.}$ (e), the Temperature-Pressure (*T-P*) phase diagram of La$_4$Ni$_3$O$_{10}$ determined by the data in Fig. 3a and 3b. The information on structure under different pressure comes from ref. [15,16]. The circular, triangle and square data dots are obtained from Fig. 3a-3b and extended data Fig. 7. There is a phase boundary between the density wave state and superconductivity, accompanied with a structure transition from monoclinic phase to tetragonal phase. (f), the temperature dependent resistance curve of monoclinic (black curve, Sample S1) and tetragonal (red curve, Sample S3) La$_4$Ni$_3$O$_{10}$ at near-ambient pressure (below 1 GPa). There is a density-wave transition evidenced by an up-turn in the resistivity at around 135 K for the monoclinic La$_4$Ni$_3$O$_{10}$, which is absent in the tetragonal La$_4$Ni$_3$O$_{10}$. (g), the temperature-dependent magnetic torque data of monoclinic and tetragonal La$_4$Ni$_3$O$_{10}$, respectively. More magnetic torque data is shown in the extended data Fig. 9. There is a kink in $\tau(T)$ at 135 K for monoclinic La$_4$Ni$_3$O$_{10}$, which is consistent with the anomaly in resistance due to density-wave transition in Fig. 1f. Similar evidences for the density-wave transition in monoclinic La$_4$Ni$_3$O$_{10}$ has been also observed by specific heats and magnetic susceptibility measurements [16]. In contrast, both resistivity and magnetic torque results indicate that such a density-wave transition is absent in the tetragonal La$_4$Ni$_3$O$_{10.}$

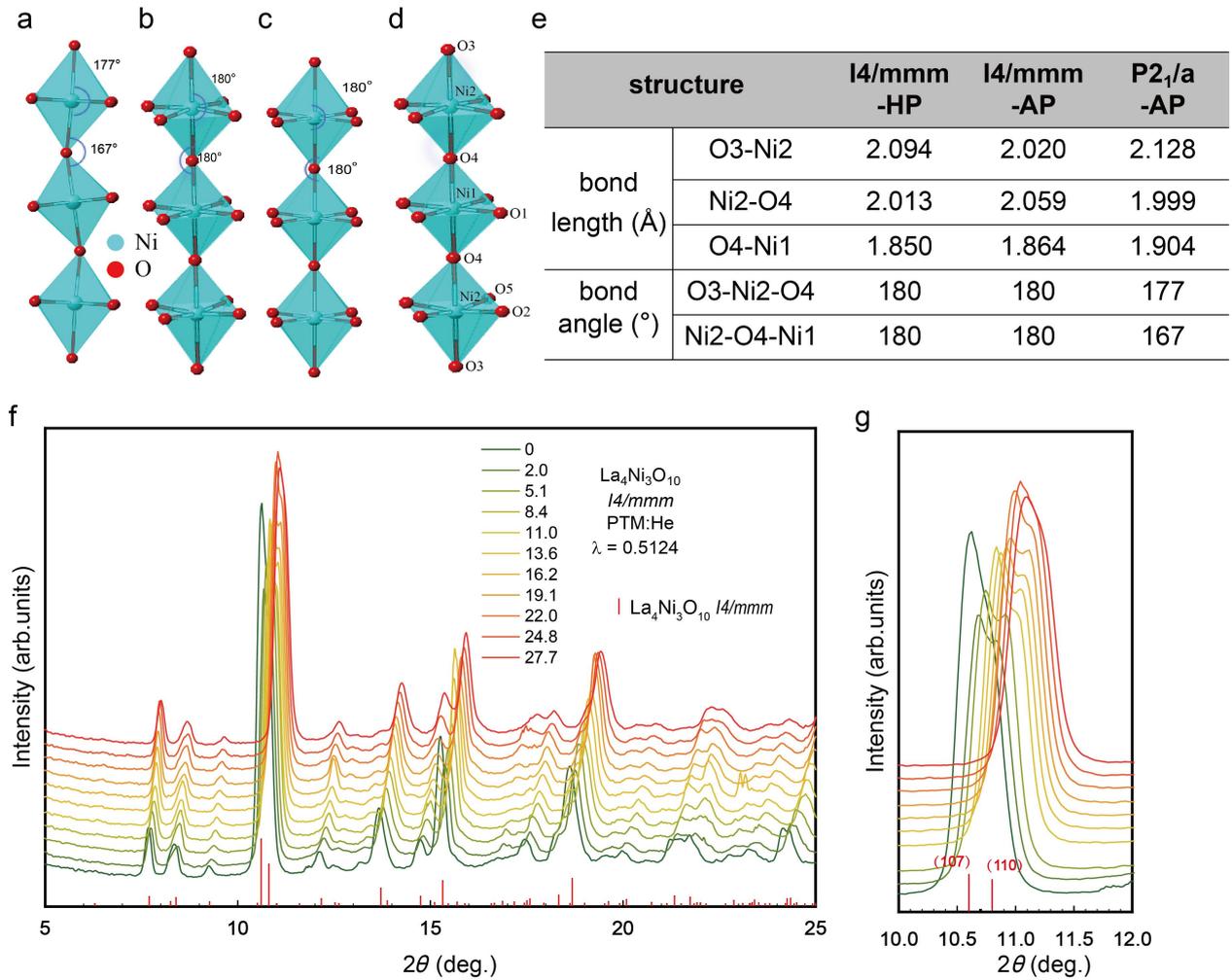

**Fig. 2. The crystal structure, bond length and bond angle comparison between different La$_4$Ni$_3$O$_{10}$.** (a)-(c), the connection between the NiO$_6$ octahedron for monoclinic phase at ambient pressure (*P2$_1$/a*-AP), tetragonal phase at 19.5 GPa (*I4/mmm*-HP) and tetragonal phase at ambient pressure (*I4/mmm*-AP), respectively. (d), the atom labels of Ni atoms and O atoms. The O5 site only exists in monoclinic phase, which becomes indistinguishable with O2 in the tetragonal phases. (e), the bond length and bond angle list between Ni atoms and O atoms. (f), the powder XRD pattern for tetragonal La$_4$Ni$_3$O$_{10}$ under the pressure range from ambient pressure to 27.7 GPa. (g), an enlarge view of (f) with 2θ between 10 - 12 degree.

**Table 1 Crystal data and structure refinement for La$_4$Ni$_3$O$_{10}$.** The structure code *I4/mmm*-AP, *P2$_1$/a*-AP and *I4/mmm*-HP represents the diffraction data for tetragonal phase at ambient pressure, monoclinic phase at ambient pressure and tetragonal phase at 19.5 GPa, respectively. The diffraction data at 19.5GPa is collected from powder XRD and the others are collected from SC-XRD. The site O5 only exists in the monoclinic phase, which merges together with site O2 in the tetragonal phase.

| structure-code | | *I4/mmm*-AP | *P2$_1$/a*-AP | *I4/mmm*-HP |
|---|---|---|---|---|
| Empirical formula | | La$_4$Ni$_3$O$_{10}$ | La$_4$Ni$_3$O$_{10}$ | La$_4$Ni$_3$O$_{10}$ |
| Formula weight | | 891.77 | 891.77 | 891.77 |
| Temperature/K | | 301.15 | 301.15 | 301.15 |
| Crystal system | | tetragonal | monoclinic | tetragonal |
| Space group | | I4/mmm | P2$_1$/a | I4/mmm |
| a/Å | | 3.8432(2) | 14.2278(7) | 3.73(1) |
| b/Å | | 3.8432(2) | 5.4655(3) | 3.73(1) |
| c/Å | | 27.942(3) | 5.4184(3) | 27.2(1) |
| α/° | | 90 | 90 | 90 |
| β/° | | 90 | 100.841(5) | 90 |
| γ/° | | 90 | 90 | 90 |
| Volume/Å$^3$ | | 412.70(6) | 413.83(4) | 378.4 |
| Z | | 2 | 2 | 2 |
| ρ$_{calc}$ g/cm$^3$ | | 7.176 | 7.157 | |
| μ/mm$^{-1}$ | | 164.05 | 163.605 | |
| F(000) | | 784 | 784 | |
| Crystal size/mm$^3$ | | 0.121 × 0.094 × 0.01 | 0.11 × 0.08 × 0.03 | |
| Radiation | | CuK$_α$ (λ = 1.54184) | CuK$_α$ (λ = 1.54184) | |
| 2Θ range for data collection/° | | 12.672 to 148.382 | 12.67 to 157.396 | |
| Index ranges | | -4 ≤ h ≤ 4, -4 ≤ k ≤ 4, -29 ≤ l ≤ 34 | -18 ≤ h ≤ 16, -6 ≤ k ≤ 6, -6 ≤ l ≤ 6 | |
| Reflections collected | | 2110 | 4920 | |
| Independent reflections | | 171 [Rint = 0.0694, Rsigma = 0.0216] | 887 [Rint = 0.0477, Rsigma = 0.0335] | |
| Data/restraints /parameters | | 171/0/20 | 887/0/73 | |
| Goodness-of-fit on F$^2$ | | 1.329 | 1.166 | |
| Final R indexes [I>=2σ(I)] | | R1 = 0.0492, wR2 = 0.1326 | R1 = 0.0454, wR2 = 0.1423 | |
| Final R indexes [all data] | | R1 = 0.0492, wR2 = 0.1326 | R1 = 0.0475, wR2 = 0.1440 | |
| Largest diff. peak/hole / e Å$^{-3}$ | | 2.01/-2.83 | 1.74/-2.93 | |
| site label | La1 | (0,0,0.4318) | (0.3644,0.4997,0.4320) | (0,0,0.4320) |
| | La2 | (0.5,0.5,0.1988) | (0.1030,0.5047,0.3013) | (0.5,0.5,0.2020) |
| | Ni1 | (0.5,0.5,0.5) | (0.5,0.5,0) | (0.5,0.5,0.5) |
| | Ni2 | (0.5,0.5,0.3606) | (0.2225,0.0014,0.3610) | (0.5,0.5,0.3580) |
| | O1 | (0.5,0,0.5) | (0.6353,0.5398,0.0705) | (0.5,0,0.5) |
| | O2 | (0,0.5,0.3607) | (0.2283,0.7498,0.1139) | (0,0.5,0.3630) |
| | O3 | (0.5,0.5,0.2841) | (0.2155,0.2497,0.6081) | (0.5,0.5,0.2810) |
| | O4 | (0.5,0.5,0.4321) | (0.0706,0.9779,0.2829) | (0.5,0.5,0.4320) |
| | O5 | | (0.5107,0.2510,0.2550) | |

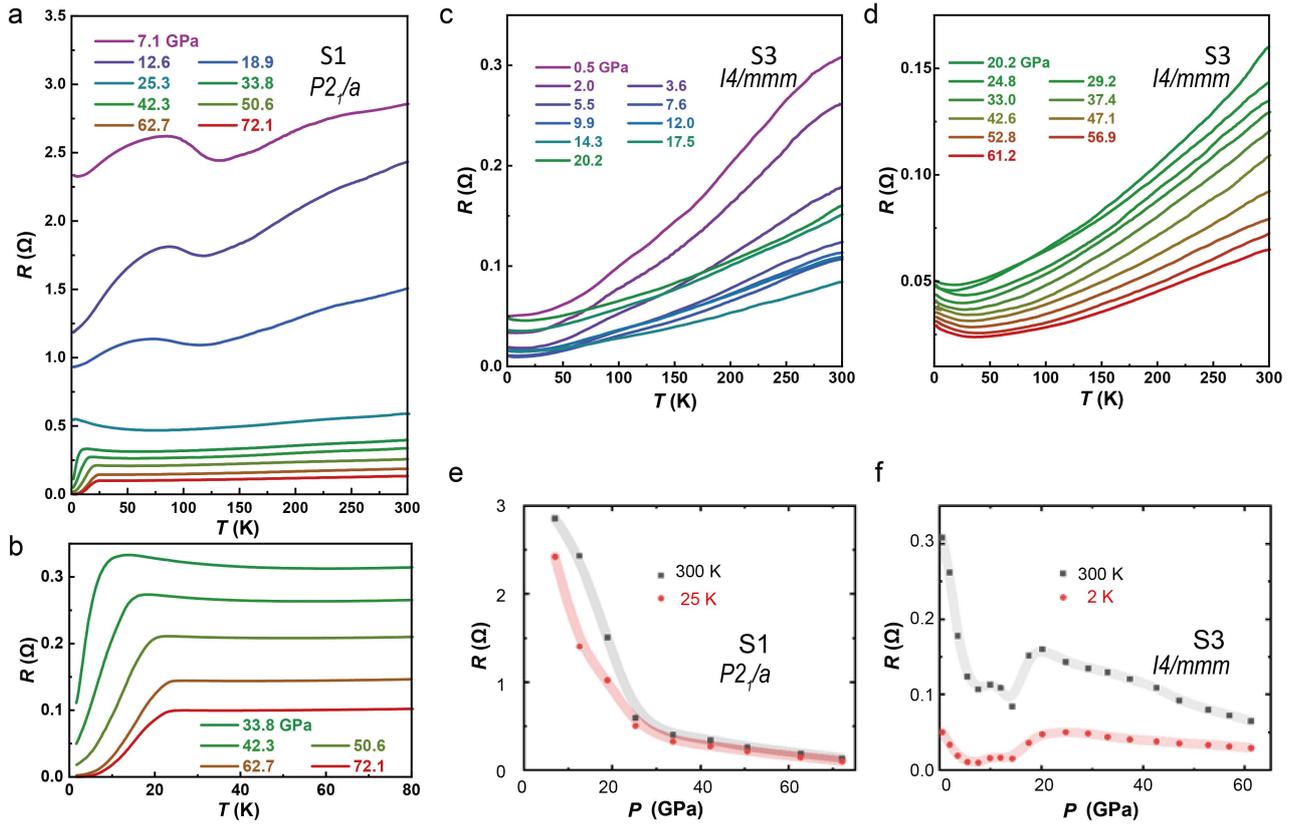

**Fig. 3. Pressure effect for $La_4Ni_3O_{10}$ with different structures.** (a), temperature dependence of resistance for monoclinic $La_4Ni_3O_{10}$ (sample S1) under various pressure. The resistance at room temperature gradually decreases with increasing pressure. The anomaly corresponding to the possible density-wave transition is gradually suppressed and superconductivity emerges above 20 GPa. With further increasing pressure, the superconducting transition becomes more pronounced and the $T_c$ continuously increases. The summarized phase diagram for density-wave and superconducting transition is show in Fig. 1e; (b), enlarged area for the resistance at low temperature which shows superconductivity. Zero resistance can be achieved above 60 GPa. (c), temperature dependence of resistance for tetragonal $La_4Ni_3O_{10}$ (sample S3) up to 20.2 GPa. The resistance gradually decreases with increasing pressure below 14 GPa, then starts to increase. No anomaly related to the density-wave transition is observed. (d), temperature dependence of resistance for tetragonal $La_4Ni_3O_{10}$ (sample S3) above 20.2 GPa. The resistance at room temperature decreases with increasing pressure and shows a weak upturn at low temperatures above 20 GPa. (e) and (f), pressure dependence of resistance at high and low temperatures for monoclinic sample S1 (e) and tetragonal sample S3 (f).

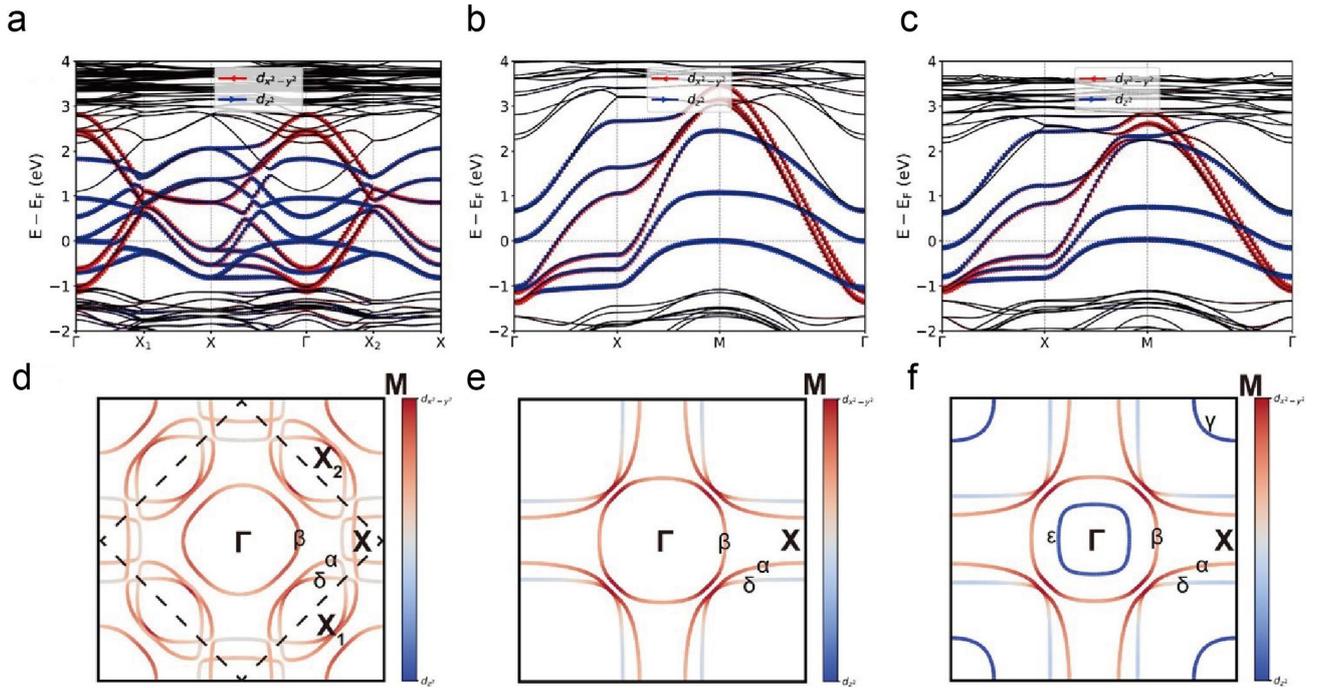

**Fig. 4. DFT calculation on La$_4$Ni$_3$O$_{10}$.** (a) The band structure of ambient pressure *P2$_1$/a* phase. (b) The band structure of high pressure (40 GPa) *I4/mmm* phase. (c) The band structure of ambient pressure *I4/mmm* phase. (d) The Fermi surface of ambient pressure *P2$_1$/a* phase. The Brillouin zone (dashed lines) is folded due to tilted octahedrons. (e) The Fermi surface of high pressure (40 GPa) *I4/mmm* phase. (f) The Fermi surface of ambient pressure *I4/mmm* phase. Notice that there are two additional Fermi pockets around Brillouin zone center and corner.

**Method**

**Sample growth:** The $La_4Ni_3O_{10}$ crystals were synthesized using a previous reported molten salt flux method [35]. The $La_2O_3$ was baked at 1000 ℃ for 10 h and then quickly transferred into the glove box with $O_2$ and $H_2O$ content less than 1ppm. The $La_2O_3$, NiO and $K_2CO_3$ were mixed with the mole ratio of 2:3:35 and loaded into a corundum crucible. The temperature was also controlled as previous reported. For monoclinic $La_4Ni_3O_{10}$, the crystals were synthesized at air. For tetragonal $La_4Ni_3O_{10}$, the crystals were synthesized under a flow oxygen. Crystals with size 70-120 um were selected from the product after dissolving the flux using water and the crystal structure were checked with the four-circle X-ray diffractometer.

**Structural characterization:** The single crystal X-ray diffraction (SC-XRD) was conducted on a four-circle diffractometer (Rigaku, XtaLAB PRO 007HF) with Cu Kα radiation in Core Facility Center for Life Sciences, USTC. The data collected at 300 K (or 110 K) were reduced and finalized using CrysAlisPro software and the structure were solved and refined using Olex-2 with ShelXT and ShelXL packages. We note that crystals grown under oxygen were firstly solved and refined with the space group *P2$_1$/a* as previous reported and results with a moderalate R1 and wR2 were obtained. After careful check with platon software and online checkcif website, a higher space group *I4/mmm* was suggested and adopted to solve and refine the crystal structure. A significant improvement in both R1 and wR2 confirm the *I4/mmm* space group. Based on the refinement from the SC-XRD data, we note that the site occupation of all the oxygen sites, especially the apical oxygen site, shows negligible vacancy which indicates the oxygen content should be very close to the ideal chemical stoichiometry. In situ high-pressure X-ray diffraction (XRD) experiments were performed using a MetalJet E1+ 160 kV source equipped with an In-Ga target, generating an X-ray wavelength of 0.5124 Å. The X-ray beam was precisely focused to an area of approximately 100×100 μm². Two-dimensional diffraction images were recorded by a PILATUS R CdTe detector and subsequently processed into one-dimensional XRD patterns using the Dioptas software. Helium gas was employed as the pressure-transmitting medium to ensure optimal hydrostatic pressure conditions, with pressure calibrations verified through the shift in the fluorescence peak of a ruby indicator. The cross-sectional transmission electron microscope (TEM) sample for selected area electron diffraction (SAED) and integrated differential phase contrast (iDPC) measurement is prepared using a scanning electron microscope equipped with a focused ion beam (Carl Zeiss Crossbeam 550L FIB-SEM). The iDPC data collected on the tetragonal $La_4Ni_3O_{10}$ sample shows no obvious oxygen vacancies or defects. The chemical composition of single crystals was characterized using a Hitachi SU8220 field emission scanning electron microscope (FE-SEM) equipped with an energy-dispersive X-ray spectroscopy (EDX, Oxford Instrument X-MaxN 150).

**Magnetic Torque characterization**: Torque magnetometry was performed using an SCL piezoresistive cantilever. The sample was attached to the tip of cantilever, which was mounted to the horizontal rotator of Physical Properties Measurement System (PPMS, Quantum Design Inc., DynaCool-14T). We first rotated the sample in a range of $\theta$ (the angle between magnetic field vector $H$ and the $c$ axis of $La_4Ni_3O_{10}$ crystal) from 0° to 90° under isothermal condition and determined that the largest signal occurs at approximately $\theta = 45°$. The data shown in Fig.1g were collected at $\theta \approx 45°$ with uncertainty of a few degrees from misalignment. We note that the measured data for this technique is the apparent resistance of the piezoresistive sensor; it indeed represents the output voltage of the Wheatstone bridge circuit inside the sensor that consists of two components, one is proportional to the magnetic torque of the sample and the other is a sample-independent background. While the latter may vary with temperature, the salient feature as shown in Fig. 1g unambiguously reflects a magnetic transition of the sample.

**High pressure transport:** Diamond anvils with various culets (100 to 300 μm) were used for high-pressure transport measurements. NaCl or Daphne oil 7373 were used as a pressure transmitting medium and the pressure was calibrated by using the shift of ruby florescence and diamond anvil Raman at room temperature. During transport measurements, the pressure was applied at room temperature using the miniature diamond anvil cell. The transport measurements were carried out in a refrigerator system (HelioxVT, Oxford Instruments). Four pieces of $La_4Ni_3O_{10}$ microcrystals are used for high-pressure transport measurement in this study. S1 and S2 are monoclinic $La_4Ni_3O_{10}$ samples. S3 and S4 are tetragonal $La_4Ni_3O_{10}$ samples.

**DFT calculations**: Our DFT calculations employ the Vienna ab-initio simulation package (VASP) code [41] with the projector augmented wave (PAW) method [42]. The meta-GGA functionals that consider the Laplacian of the electron density [43] are used here. Typically, this can be more accurate compared to the GGA method. The cutoff energy for expanding the wave functions into a plane-wave basis is set to be 500 eV. The energy convergence criterion is $10^{-8}$ eV. All calculations are conducted using the primitive cell to save time. The Γ-centered 9×9×4, 11×11×15 and 11×11×15 k-meshes are used for the ambient pressure *P2₁/a* phase, ambient pressure *I4/mmm* phase and high pressure I4mmm phase respectively. In order to determine the magnetic state, we set three different magnetic states which are usually encountered in this type of materials: ferromagnetic (FM), G-type antiferromagnetic (G-AFM) ordering at (*π, π*) and double stripe ordering at (*π/2, π/2*). We compare them with the non-magnetic (NM) state. At all U values shown here, we find the energies of FM state are relatively lower than other two competing magnetic states. We also calculate the magnetic moment of three magnetic states. In two AFM states (G-AFM and double stripe), the middle layer has zero magnetic moment, and the sign of magnetic moment is opposite in top and bottom layer. We calculate the spin orders using the simplified rotation invariant approach to the DFT+U, introduced by Dudarev et al [44].

## Acknowledgments

This work is supported by the National Key R&D Program of the MOST of China (Grant No. 2022YFA1602601), the National Natural Science Foundation of China (Grants No. 11888101, 12034004, 12161160316, 12325403), the Strategic Priority Research Program of Chinese Academy of Sciences (Grant No. XDB25000000), the Chinese Academy of Sciences under contract No. JZHKYPT-2021-08, the CAS Project for Young Scientists in Basic Research (Grant No. YBR-048), and the Innovation Program for Quantum Science and Technology (Grant No. 2021ZD0302800).

## Author contributions

X.H.C. conceived the research project. X. H. and T. W. coordinated the experiments. M.Z.S. grew the single crystals and performed the structural characterizations. Y.K.L. and J.J.Y. performed the high-pressure transport measurement. Y.X.W. and K.J performed the DFT calculations. D. P. and Q.S. Z. measured the powder XRD data under different pressures. S.H. Y., D.S. S. and B.H. G. conducted the TEM measurement. H.P. L. and K.B. F. measured the magnetic torque data. M.Z.S., J.J.Y., T.W. and X.H.C. analyzed the data with the help from K.J., J.F.H, Z.J.X., Z.Y.W. M.Z.S., J.J.Y., K.J., T.W. and X.H.C. wrote the paper with inputs from all authors.